\documentclass{article}
\usepackage{spconf,amsmath,graphicx}

\usepackage{enumitem}
\setlist{nosep, leftmargin=14pt}

\usepackage{mwe} 

\title{Self Supervised Lesion Recognition for Breast Ultrasound Diagnosis}
\name{Yuanfan Guo$^{1,3}$, Canqian Yang$^{1,3}$, Tiancheng Lin$^{1,3}$, Chunxiao Li$^{2}$, Rui Zhang$^{1,3}$, Rong Wu$^{2}$, Yi Xu$^{1,3}$\sthanks{Corresponding Author}}
\address{$^{1}$Shanghai Key Lab of Digital Media Processing and Transmission, Shanghai Jiao Tong University \\  $^{2}$Department of Ultrasound, Shanghai General Hospital, Shanghai Jiao Tong University School of Medicine \\$^{3}$MoE Key Lab of Artificial Intelligence, AI Institute, Shanghai Jiao Tong University}
\begin{document}
%
\maketitle
\begin{abstract}

Previous deep learning based Computer Aided Diagnosis
(CAD) system treats multiple views of the same lesion as independent
images. Since an ultrasound image only describes a partial 2D projection of a 3D lesion, such paradigm ignores the semantic relationship between different views of a lesion, which is inconsistent with the traditional diagnosis where sonographers analyze a lesion from at least two views. In
this paper, we propose a multi-task framework that complements Benign/Malignant classification task with lesion recognition (LR) which
helps leveraging relationship among multiple views of a single lesion to
learn a complete representation of the lesion. To be specific,
LR task employs contrastive learning to encourage representation that
pulls multiple views of the same lesion and repels those of different lesions. The task therefore facilitates a representation that is not only invariant to the view change of the lesion, but also capturing fine-grained features to distinguish between different lesions. Experiments show that the proposed multi-task framework
boosts the performance of Benign/Malignant classification as two sub-tasks complement each other and enhance the learned representation of
ultrasound images.

\end{abstract}

\section{Introduction}
Breast cancer is one of the leading cause of death for women worldwide. With the advantage of being non-invasive, safe, and relatively inexpensive~\cite{ultrasound_advantage}, Breast Ultrasound (BUS) is a widely adopted imaging modality for early breast cancer diagnosis. An ultrasound image is a 2D projection of a 3D lesion describing only partial characteristics of the lesion. Generally, radiologists diagnose a lesion by referring to multiple ultrasound images of different views since a singular view of the lesion may show ambiguous morphological characteristics that contributes little to Benign/Malignant discrimination~\cite{ultrasound_multiviews}. 
To reduce workload of radiologists and improve diagnostic accuracy, deep learning based computer-aided diagnosis (CAD) system has been developed to help radiologists in breast cancer Benign/Malignant classification~\cite{ultrasound_BM1,ultrasound_BM2}. However, in these works, different views of the same lesion are considered as independent images, which ignores the complementary information among different views and is inconsistent with the diagnosis mechanism of radiologists.

Intuitively, a deep learning based CAD system for lesion diagnosis should also leverage all possible ultrasound views of the same lesion for final decision making. In other word, a classification model is expected to learn representations that are invariant to the view change of the same lesion, which implicitly encourages the model to discriminate between different lesions. This ability of the model as a lesion-level discrimination is appealing since it is supposed that the common features to identify individual sample may also capture similarity among samples that facilitates semantic-level classification. To this end, we are interested in instance recognition (IR)~\cite{IR,MoCo}, a typical contrastive learning task that learns representations invariant to data augmentation on images. 
This is achieved by pulling together two randomly augmented versions of the same image and repelling apart those of different images in the embedding space. IR has been proved to be effective as a self-supervised pretext task for different downstream tasks including image classification, object detection and segmentation for both natural images~\cite{IR} and medical images~\cite{comparing_to_learn,azizi2021big}. 
Besides, recent studies have presented theoretically and empirically ~\cite{ruder2017overview} that shared representation among related tasks in a multi-task framework facilitates the generalization of deep learning model, where IR usually serves as one of the useful auxiliary tasks to improve the main task.



Inspired by works mentioned above and based on the insight of exploiting multi-views information, we propose lesion recognition (LR) as an auxiliary task in a multi-task framework to complement the Benign/Malignant classification. 
Instead of identifying each individual image as done in IR, LR enforces similarity among images of the same lesion and allows variance across images from different lesions. Therefore, LR encourages the model to identify which lesion a ultrasound image corresponds to.
It should be noted that the proposed LR task requires no additional annotation, because ultrasound image data are naturally organized such that images of the same lesion are stored together. The benefits of LR come from two aspects. 
On the one hand, it reduces the inconsistency among predictions of different images from the same lesion, which helps integrating information from all possible views of the same lesion and thus makes a more accurate diagnosis. 
On the other hand, learning to distinguish different lesions requires capturing lesion-specified representation, which are probably more fine-grained than the category-specified (Benign/Malignant) counterparts. This might lead to diversified and enhanced features that help boost the performance of semantic classification.
The main contribution of this paper could be summarized as:
\begin{itemize}
    \item We propose to leverage multi-view information that naturally embedded in breast ultrasound data to enhance the learned representation of lesions.
    \item We propose a multi-task framework, where a novel lesion recognition task is employed to complement and improve the main classification task.
    \item We achieve an improvement of $1.5\%$ in area under the receiver operating characteristic curve (AUC) and $2.8\%$ in accuracy on main classification task with the proposed LR task employed.
\end{itemize}


\begin{table*}[ht]
    \centering
    \setlength{\belowcaptionskip}{-20pt}
    \caption{Benign/malignant classification performance (mean$\pm$std\%) with different auxiliary tasks introduced.}
    \begin{tabular}{c|c|c|c|c|c|c|c}
        \hline
        Method & AUC  & ACC & Sensitivity  & Precision  & Specificity  & F1  & MCR $\downarrow$ \\
        \hline
        Baseline & 94.7$\pm$1.7 & 86.3$\pm$1.7 & 92.6$\pm$3.8 & 87.6$\pm$2.6 & 76.3$\pm$5.1 & 89.9$\pm$1.4 & 59.9$\pm$4.6 \\
        \hline
        IR & 93.9$\pm$1.7 & 84.5$\pm$2.6 & 86.6$\pm$8.6 & 90.3$\pm$5.6 & 80.8$\pm$12.6 & 88.0$\pm$2.6 & 63.3$\pm$7.9 \\
        \hline
        LR  & \textbf{96.2$\pm$1.0} & \textbf{89.1$\pm$1.9} & 92.3$\pm$3.7 & \textbf{91.5$\pm$2.4} & \textbf{83.1$\pm$5.5} & \textbf{91.8$\pm$1.4} &
        \textbf{52.7$\pm$4.1}\\
        \hline
    \end{tabular}
    \label{MainExperiment}
\end{table*}

\section{Method}
\subsection{Overall Framework}
\begin{figure}[tb]
\setlength{\abovecaptionskip}{-5pt}
\setlength{\belowcaptionskip}{-5pt}
\includegraphics[width=0.5\textwidth]{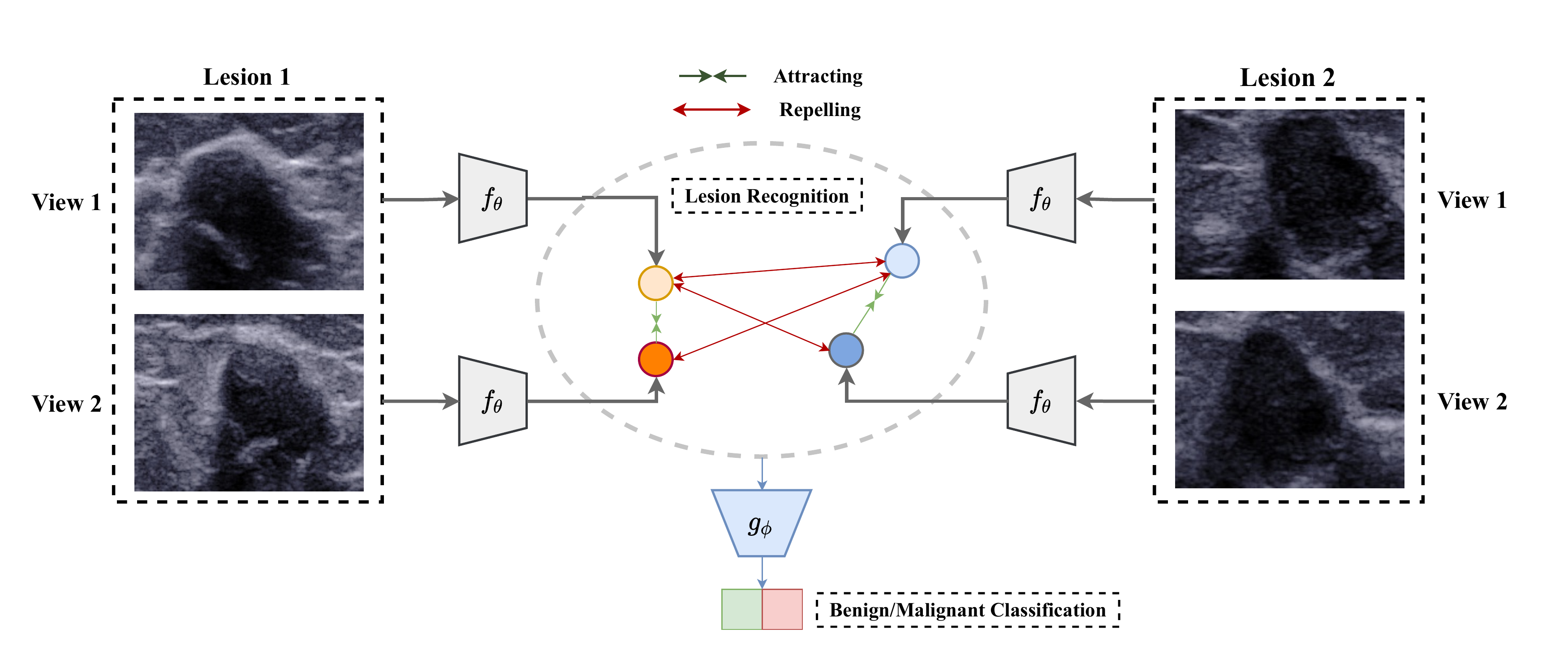}
\caption{Overview of the proposed multi-task framework. Given a set of different ultrasound images containing views of different lesions, a CNN as backbone network $f_\theta(\cdot)$ is applied to extract features, followed by a contrastive loss to pull together features of images from the same lesion and repel apart those from different lesions. After that, the extracted features are fed into classification head (fully-connected layer) $g_\phi(\cdot)$ to perform Benign/Malignant prediction, which is guided by classification loss.
} 
\label{OverallFramework}
\end{figure}

Figure~\ref{OverallFramework} illustrates the overall framework of the proposed method. It mainly consists of: (1)  feature extraction from ultrasound images through a convolutional neural network (CNN) $f_\theta(\cdot)$, (2) Benign/Malignant classification as the main task that is achieved by a classification head $g_\theta(\cdot)$, instantiated as a fully-connected layer, (3) Lesion Recognition as an auxiliary task to enhance the features extracted by backbone network.

\subsection{Main Task: Benign/Malignant Classification}

Benign/Malignant classification is the main task in the proposed multi-task framework to perform lesion diagnosis. Given a batch of $N_b$ ultrasound images and corresponding labels $\mathcal{B}=\{x_i, y_i\}_{i=1}^{N_b}$, where $y_i\in\{0,1\}$ (Benign/Malignant), the purpose of the main task is to learn a CNN backbone network $f_\theta(\cdot)$ parameterized by $\theta$ and a classification head network $g_\phi(\cdot)$ parameterized with $\phi$ that automatically predict the probability of certain image $x_i$ being a view of a Benign/Malignant lesion. This can be achieved by optimizing the common cross entropy loss:
\begin{equation}
    \mathcal{L}_{cls} = - \frac{1}{N_b} \sum_{i=1}^{N_b} y_i \log( g_\phi(f_\theta(\mathcal{T}(x_i))),
\end{equation}
where $\mathcal{T}(\cdot)$ is a random data augmentation such as flipping and random cropping during training.
\subsection{Auxiliary Task: Lesion Recognition}

As mentioned in the previous section, multiple images representing a lesion from different views provide complementary information for breast ultrasound diagnosis. With such information  naturally available in ultrasound data, we propose to employ Lesion Recognition (LR) as an auxiliary task. To be specific, we hope to pull features representing the same lesion together and push those representing different lesions far apart. The task is expected to facilitate the fine-grained representations that contribute to the main task with the multi-view complementary information. This could be achieved by minimizing a contrastive loss, which is generally formulated as:
\begin{equation}
    \mathcal{L}_{con} = - \frac{1}{N_b} \sum_{i=1}^{N_b} \sum_{j\in \mathcal{P}(i)} \log \frac{\exp(S(z_i,z_j))}{\sum_{k\in \mathcal{N}(i) \cup \mathcal{P}(i)} \exp(S(z_i,z_k))},
\end{equation}
where $z_i=f_\theta(\mathcal{T}(x_i))$ denotes feature vector extracted by the backbone network. $\mathcal{P}(i)$ and $\mathcal{N}(i)$ are the sets of indices of samples that are defined as positive and negative respectively to the anchor sample $x_i$.
$S(\cdot, \cdot)$ is similarity function, which is implemented as cosine similarity in this paper. It could be seen that minimizing the contrastive loss increases the similarity of positive pairs $(z_i,z_j)$ and decreases that of negative pairs. Recall the fact that images of the same lesion are actually different views of an identical object, for our proposed LR task, positive pairs are defined as images from same lesion whereas negative pairs are images from different lesions: 
\begin{equation}
    \mathcal{P}(i)=\{j|L(j)=L(i)\},
    \mathcal{N}(i)=\{k|L(k)\neq L(i)\},
\end{equation}
where $L(i)$ denotes the lesion index of image $x_i$. Such definition of positive and negative pairs brings two potential advantages. On the one hand, information contained in different views of a lesion promote a complete representation of the lesion. This in turn benefits each view of the lesion in contributing to a more accurate diagnosis by reducing the inconsistency among predictions of different views. On the other hand, the model is required to capture find-grained information, which enhances the representation learned by model.

Finally, the overall framework is optimized by a joint loss as:
\begin{equation}
    \mathcal{L} = \mathcal{L}_{cls} + \alpha \mathcal{L}_{con},
\end{equation}
where $\alpha$ is weighting parameter.

\section{Experiments}
\subsection{Data and Implementation Details}
The proposed method is evaluated on an in-house breast ultrasound dataset. The collected dataset contains a total of $5,911$ images from $2,131$ lesions, where $2,002$ images are benign and $3,909$ are malignant. $5$-fold cross-validation is applied on the dataset to assess the proposed method and all comparison counterparts. Specifically, the dataset is randomly split into $5$ disjoint folds and the validation consists of $5$ rounds. At each round, the method is trained on $4$ folds and evaluated on the holdout fold.

To validate the proposed method, the following counterparts are included: (a) \textbf{baseline}: the network with only $\mathcal{L}_{cls}$; (b) \textbf{Lesion Recognition (LR)}: the network with joint loss formulated by Eq(3)(4); (c) \textbf{Instance Recognition (IR}): the network with joint loss formulated by Eq(4)(5).

ResNet-50~\cite{ResNet} is adopted as the backbone network. All mini-batches are constructed by arbitrarily sampling $8$ patients and $8$ images per patient, resulting in a mini-batch size of $64$. As for image augmentation, all images are resized into $256\times 256$ pixels, followed by random cropping, flipping and color jittering to prevent overfitting. Adam optimizer \cite{kingma2014adam} is employed to train all methods with an initial learning rate of $1e-4$. The learning rate is decayed by a factor of $0.1$ every $50$ epochs and it takes $200$ epochs for the models to converge. All experiments were conducted on an NVIDIA TITAN X GPU.

\subsection{Benign/Malignant Classification}
The comparison of Benign/Malignant classification performance is presented in Table~\ref{MainExperiment}. The result shows that employing LR as an auxiliary task achieves an improvement of $1.5\%$ in AUC and $2.8\%$ in ACC compared to baseline. On the contrary, employing IR degrades the performance by $0.8\%$ in AUC and $1.8\%$ in ACC, 
illustrating the necessity of proper definition of positive and negative pairs. It should be noted that with LR introduced, the specificity and precision increase significantly by $6.8\%$ and $3.9\%$, respectively, indicating that the involvement of LR task reduces the bias of model towards positive predictions.


\begin{table*}[tb]
    \centering
    \setlength{\belowcaptionskip}{-20pt}
    \caption{Ablation study on negative samples with LR employed on classification task(mean$\pm$std\%). `LR(-SC)' denotes removing negative samples from same semantic class. `LR(-DC)' denotes removing negative samples from different semantic class. `LR(-)' denotes removing all negative samples.}
    \begin{tabular}{c|c|c|c|c|c|c}
        \hline
        Method & AUC & ACC & Sensitivity & Precision & Specificity & F1 \\
        \hline
        LR(-) & 94.9$\pm$1.3 & 86.1$\pm$1.2 & 91.4$\pm$4.2 & 88.2$\pm$2.9 & 76.0$\pm$6.2 & 89.7$\pm$1.0  \\
        \hline
        LR(-SC) & 95.0$\pm$1.4 & 86.6$\pm$1.2 & 91.9$\pm$3.1 & 88.4$\pm$1.9 & 76.5$\pm$4.1 & 90.1$\pm$1.0 \\
        \hline
        LR(-DC) & 95.6$\pm$1.1 & 88.3$\pm$1.8 & 91.6$\pm$2.9 & 90.9$\pm$1.7 & 81.9$\pm$3.9 & 91.2$\pm$1.3 \\
        \hline
        LR  & 96.2$\pm$1.0 & 89.1$\pm$1.9 & 92.3$\pm$3.7 & 91.5$\pm$2.4 & 83.1$\pm$5.5 & 91.8$\pm$1.4 \\
        \hline
    \end{tabular}
    \label{AblationDifficulty}
\end{table*}

To better demonstrate the effectiveness of the proposed LR for reducing inconsistent predictions among multiple views of the same lesion, 
inner lesion mis-classification rate for a single lesion $L_i$, denoted as $MCR(L_i)=W_{L_i}/N_{L_i}$,
where $W_{L_i}$ is the number of mis-classified images and $N_{L_i}$ is the total number of images contained in $L_i$, is employed as an extra performance measure. Specifically, the average $MCR$ among lesions containing at least $1$ mis-classified image is reported in Table~\ref{MainExperiment}. As could be seen, compared to the baseline, employing LR reduces $MCR$ by $7.2\%$, while employing IR increases $MCR$ by $3.4\%$, indicating a severer inner lesion inconsistency as IR introduced. This is because that the objective of IR encourages a view-variant representation of lesion as discussed in section 2.3, which is critically contradict to the objective of the main task.

As investigated in~\cite{frankle2020all}, negative samples play a significant role in contrastive learning as the learning will be affected by the difficulty of negative samples.
For the proposed LR, negative samples in the same class are more difficult than those in different class, since the negatives in the same class may share some similarity thus it is hard for LR to discriminate between them. To investigate sufficiency and necessity of different negative samples, we conduct ablation experiments by removing different sets of negative samples. As could be seen in Table~\ref{AblationDifficulty}, 
removing all negative samples degrades the AUC significantly by $1.1\%$, indicating that negative samples are indeed necessary for LR task. 
Further, performance gap between `LR(-SC)' and `LR(-DC)' ($95.0\%$ vs. $95.6\%$ in AUC) demonstrates that hard negatives are more sufficient than easy ones, which is consistent with the conclusion in~\cite{frankle2020all}. 
Finally, the result of using all negative samples suggests that large number of negative samples to compare is the guarantee of performance.

\section{Conclusion}
Radiologists will refer to multiple views of a breast lesion for diagnosis. 
In this paper, we propose to exploit the lesion-image structure of ultrasound data where multiple views of a lesion are available. 
A self-supervised contrastive learning task LR is proposed to enhance the performance of Benign/Malignant classification by encouraging  the model to learn representation that is robust to change of viewpoint and distinguishable between lesions. The experimental results show that the involvement of LR as an auxiliary task boosts the performance of semantic classification.
We hope that this work could help utilizing the free annotations embedded in the organization of ultrasound data, which shows a potential to enhance the performance of semantic classification task.

\textbf{Compliance with Ethical Standards}:
All procedures performed in studies involving human participants were in accordance with the ethical standards of the institutional and/or national research committee.

\textbf{Acknowledgments}:
This work was supported in part by National Natural Science Foundation of China 62171282, 111 project BP0719010, Shanghai Jiao Tong University Science and Technology Innovation Special Fund ZH2018ZDA17, and Shanghai Municipal Science and Technology Major Project (2021SHZDZX0102).



\bibliographystyle{IEEEbib}
\bibliography{strings,refs}

\end{document}